\documentclass[conference]{IEEEtran}
\IEEEoverridecommandlockouts
\usepackage{algorithm,algorithmic}
\usepackage{cite}
\usepackage{tabularx}
\usepackage{makecell}
\usepackage{mdframed}
\usepackage[table]{xcolor}
\usepackage{adjustbox}
\usepackage{adjustbox}
\usepackage{array}
\usepackage{makecell}
\usepackage{multirow}
\usepackage{fancyhdr}
\usepackage{amsmath,amssymb,amsfonts}
\usepackage{algorithmic}
\usepackage{graphicx}
\usepackage{textcomp}
\usepackage{xcolor}
\usepackage{hyperref}
\usepackage{float}
\usepackage{fancyhdr}
\usepackage{lastpage}
\usepackage{eso-pic}
\usepackage{fancyhdr}
\usepackage{tikz}
\usepackage{amsmath}
\usepackage{subcaption}
\usepackage{multicol}
\usepackage{balance}
\usepackage{float}
\usepackage{xspace}
\usepackage{array}
\usepackage{multirow}
\usepackage{array} 
\usepackage{graphicx}
\usepackage{subcaption}
\usepackage{xspace}
\usepackage{mdframed}
\usepackage{multirow}
\usepackage{adjustbox}
\usepackage{breakurl}
\usepackage{url}
\usepackage{tabularx}
\usepackage{hyperref}
\usepackage{siunitx}
\usepackage{xurl}
\newcommand{\name}{\textit{BugReAct}}
\newcommand{\react}{\textit{ReAct}}
\usepackage{caption} 
\usepackage{caption} 
\def\BibTeX{{\rm B\kern-.05em{\sc i\kern-.025em b}\kern-.08em
    T\kern-.1667em\lower.7ex\hbox{E}\kern-.125emX}}

\fancypagestyle{FirstPage}{
\lfoot{979-8-3315-1909-4/24/\$31.00 \copyright2024 IEEE} 
}

\makeatletter

\def\ps@IEEEtitlepagestyle{%
    \def\@oddfoot{\mycopyrightnotice}%
    \def\@evenfoot{}%
}
\def\mycopyrightnotice{%
    {\footnotesize  979-8-3315-1909-4/24/\$31.00 \textcopyright2024 IEEE\hfill}
    \gdef\mycopyrightnotice{}
}
\makeatother

\usepackage{eso-pic}

\begin{document}

\title{When Agents Fail: A Comprehensive Study of Bugs in LLM Agents with
Automated Labeling}

\author{ \IEEEauthorblockN{ Niful Islam\textsuperscript{1}, Ragib Shahariar Ayon\textsuperscript{2}, Deepak George Thomas\textsuperscript{3}, Shibbir Ahmed\textsuperscript{2}, Mohammad Wardat\textsuperscript{1,*} } \IEEEauthorblockA{ \textsuperscript{1}Oakland University, Rochester Hills, USA\\ \textsuperscript{2}Texas State University, San Marcos, USA\\ \textsuperscript{3}Tulane University, New Orleans, USA } }
\maketitle
\thispagestyle{plain}
\pagestyle{plain}
\let\thefootnote\relax\footnotetext{*Corresponding author: wardat@oakland.edu}

\begin{abstract}
Large Language Models (LLMs) have revolutionized intelligent application development. While standalone LLMs cannot perform any actions, LLM agents address the limitation by integrating tools. However, debugging LLM agents is difficult and costly, as the field is still in its early stages and the community is underdeveloped. To understand the bugs encountered during agent development, we present the first comprehensive study of bug types, root causes, and effects in LLM agent-based software. We collected and analyzed 1,268 bug-related posts and code snippets from Stack Overflow, GitHub, and Hugging Face forums, focused on LLM agents built with seven widely used LLM frameworks as well as custom implementations. For a deeper analysis, we have also studied the component of the LLM agent where the bug occurred, along with the programming language and framework. This study also investigates the feasibility of automating bug identification. For that, we have built a~\react~agent named~\name, equipped with adequate external tools to determine whether it can detect and annotate the bugs in our dataset. According to our study, we found that~\name~equipped with Gemini 2.5 Flash achieved a remarkable performance in annotating bug characteristics with an average cost of 0.01 USD per post/code snippet.
\end{abstract}


\section{Introduction}

Large Language Model (LLM) agents are autonomous systems with human-like abilities such as autonomy, reasoning, and planning~\cite{huang2025genesis}. With recent advancements in LLMs, their adoption has grown rapidly across safety-critical domains including security~\cite{huang2025ai_sec}, healthcare~\cite{huang2025ai_health}, robotics~\cite{Huang2025_robots}, and end-to-end software development~\cite{liu2024large, hosseini2025role}. Like any software, LLM-based agents are susceptible to bugs, and understanding their types, root causes, and effects is essential for the software engineering community to mitigate them, a challenge compounded by the rapid proliferation of LLM agent frameworks.

While LLM agents are a form of software, they differ from traditional systems in three key ways. First, their reliance on black-box LLMs makes debugging more complex~\cite{black-box}. Second, they are nondeterministic, meaning the same input can produce different outputs depending on temperature settings~\cite{ouyang2025empirical}, which makes bug reproduction more challenging. Third, the LLM agent ecosystem is still nascent. For example, LangChain, the most widely used framework, was only released in late 2022, and fewer experienced developers are available on community platforms to provide support~\cite{awesome-llm-agents}. These factors collectively motivate an extensive study of bugs in LLM agents.

Recent work has examined defects in LLM agents and proposed taxonomies of agent-based bugs and errors~\cite{ning2024defining, rahardja2025can}. However, one effort focuses only on defect types, while the other primarily analyzes bugs in agent-building frameworks rather than the agents themselves. Additionally, while LLM agents have demonstrated exceptional abilities across multiple fields~\cite{zhujudgelm, li2024llms}, their effectiveness drops significantly when tasked with fixing issues within agentic systems. For example, the best-performing agent achieved only 4\% accuracy~\cite{rahardja2025can}, which highlights their currently limited self-healing capability.

To fill this gap, we present the first study of bugs in LLM agents across multiple platforms. Following Islam~\textit{et al.}~\cite{islam2019comprehensive}, we characterize bugs by type, root cause, and effect, and identify the most affected agent component. We also analyze the distribution of bug types, root causes, and frameworks, and examine how bugs evolved over time. Additionally, we built a~\name~agent that automatically labels bugs from online sources to assist developers in resolving them.

Building this study presented several challenges. Given the recent emergence of LLM agents, developers tend to seek help through conversational AI tools rather than conventional platforms, which reduces the quantity and quality of data available in traditional sources~\cite{humbatova2020taxonomy, islam2019comprehensive}. Therefore, in addition to Stack Overflow and GitHub, we extended our search to Hugging Face forums for broader data collection. To mitigate biases from foundation models,~\name~was developed using three distinct models, which prevents findings from depending on any single one. Finally, since the field evolves rapidly, encountered bugs were often novel, which posed a non-trivial challenge for both human and automated annotators. We release the dataset to assist developers in navigating this dynamic field. In summary, this paper makes the following main contributions:

 \begin{itemize}
     \item To the best of our knowledge, we conduct the first large-scale empirical study on LLM agent bugs. We collect bugs from Stack Overflow, GitHub, and Hugging Face across seven frameworks and organize into a taxonomy.
     \item We propose~\name, an LLM-based agent that automatically annotates bug-related posts by analyzing content, and invoking external tools.
     \item We evaluate \name on 1,268 posts across three LLMs. It achieves bug characterization comparable to manual labeling and shows potential to support bug repair.
     
 \end{itemize}

\section{Related Work}
\label{sec:background}

\paragraph{\textbf{Empirical Study on Bugs}}

There has been extensive research on bugs in data mining and information retrieval tools~\cite{thung2012empirical}, deep learning systems~\cite{humbatova2020taxonomy, islam2019comprehensive}, attention-based models~\cite{jahan2025taxonomy}, deep reinforcement learning~\cite{thomas2025mu}, and LLM libraries~\cite{jiang2025foundation, han2025comprehensive}, with studies focused on root causes, fault taxonomies, and bug patterns. While limited prior work has focused on LLM agents specifically, Xue~\textit{et al.}~\cite{characterizationXue} and Zhang~\textit{et al.}~\cite{zhang2026dissecting} analyzed bugs across LLM agent frameworks. Pan~\textit{et al.}~\cite{pan2025multiagent} analyzed execution traces to investigate failures in Multi-Agent Systems (MAS). Islam \textit{et al.} \cite{Islam2026SelfHealEF} conducted a study analyzing bug-fix patterns in LLM agents. Lastly, Rahardja \textit{et al.}~\cite{rahardja2025can} built a taxonomy of LLM agent issues from 201 GitHub issues across 18 repositories, 11 of which are agent frameworks such as CrewAI, and evaluated SoTA coding agents on agentic tasks: Agentless~\cite{xia2025demystifying}, AutoCodeRover~\cite{zhang2024autocoderover}, and SWE-agent~\cite{yang2024swe} resolved only 3.33\%, 1.33\%, and 0.67\% of agentic issues, versus 32.0\%, 30.67\%, and 18.33\% on traditional tasks. Their taxonomy is built largely from framework and tool issues, while ours focuses on bugs that originate from the agents themselves.

\paragraph{\textbf{Agentic AI for SE and SE for Agentic AI}}
In existing literature, agents have been used as a tool for fixing errors in software systems. For instance, Agentless~\cite{xia2025demystifying}, an LLM agent, is used for bug localization, repair, and patch validation. Similarly, AGDebugger~\cite{epperson2025interactive} addresses multi-agent debugging through interactive features and real-time agent steering. RAFFLES~\cite{zhu-etal-2026-raffles} is used as an iterative Judge-Evaluator framework for automated fault attribution. Nonetheless, some research has evaluated bugs in agentic systems as well. Ning~\textit{et al.}~\cite{ning2024defining} mined Stack Overflow posts on LLM agents and developed a static analysis approach for defect detection. Unlike their work, which focuses solely on defect types, our study examines bug types, root causes, effects, and responsible components, and relies on a~\name~agent for automatic annotation. Deshpande~\textit{et al.}~\cite{deshpande2025trail} proposed a taxonomy of agent errors and investigated whether LLMs could diagnose them using agent traces, though this requires access to the complete codebase and consumes considerable tokens. Zhu~\textit{et al.}~\cite{zhu2025llm} constructed AgentErrorBench, a dataset of 200 annotated failure trajectories, and proposed AgentDebug, a three-stage framework that traces failures to their root causes. Cemri~\textit{et al.}~\cite{pan2025multiagent} were the first to investigate failure modes in Multi-Agent Systems, and proposed a taxonomy alongside an LLM trained to identify them from benchmark traces~\cite{jimenez2024swe}.

Overall, existing studies either examine bugs across broad domains or focus on agent-building frameworks, leaving a gap in understanding bugs in agentic systems, which persists due to their black-box nature, dynamic behavior, and nascent community. This work addresses both gaps through an extensive study of bugs in LLM agents, with a focus on bug characteristics and practical insights for developers, library maintainers, and researchers, and further evaluates the ability of LLM agents to annotate bugs in their own systems through \name, a ReAct-based agent built for this purpose.

\section{Empirical Study}
\label{sec:empirical_study}
\begin{figure*}[h]
    \centering
\includegraphics[height=8cm,width=1\linewidth]{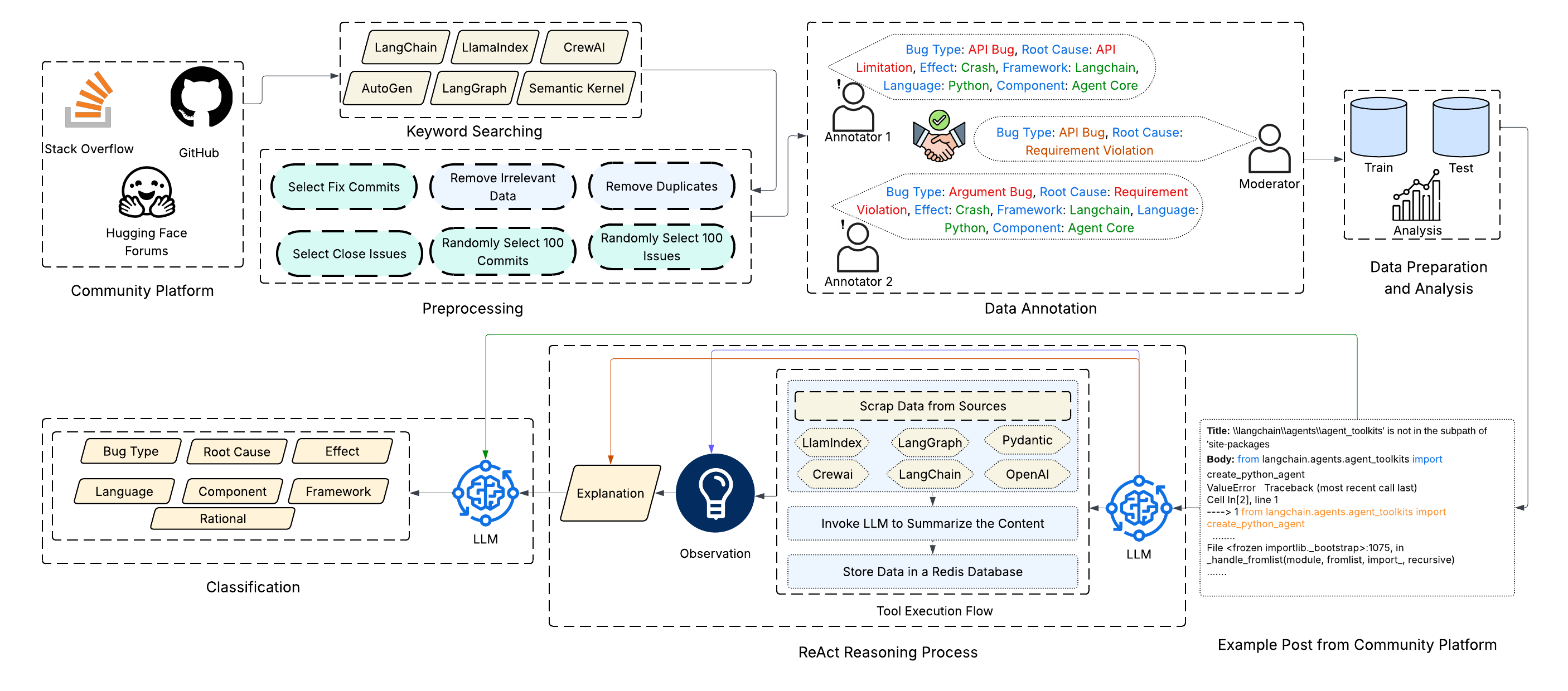}
    \caption{Workflow of the overall approach.}
    \label{fig:workflow}
\end{figure*}

As presented in Figure \ref{fig:workflow}, the proposed approach consists of two main steps: collecting and annotating data for the empirical study, and building an LLM agent to evaluate automated bug classification. This section discusses the empirical study conducted to assess bugs encountered by developers during the design of LLM agents. The research team manually analyzed 4,341 bug-related posts and code snippets, from which 1,268 instances were annotated for the study. The empirical study addresses the following research questions. 

\begin{itemize}
    \item \textbf{RQ1 (Bug Characterization):} What are the most common bug types, root causes, and effects in developing LLM agents? 
    \item \textbf{RQ2 (Bug-Prone Components):} Which component of the LLM agent is more prone to bugs? 
    \item \textbf{RQ3 (Bug Mapping):} What is the distribution across bug types, root causes, and frameworks used?
    \item \textbf{RQ4 (Bug Evolution): }How do different types of bugs evolve in frequency over time?
\end{itemize}
\vspace{-7pt}

\subsection{Dataset}
For the empirical study, we collected a large dataset and annotated it based on bug type, root cause, effect, the LLM agent component involved in bug occurrence, programming language, and development framework. Following previous works \cite{islam2019comprehensive, humbatova2020taxonomy}, two sources were employed in this study, namely the Stack Overflow forum and GitHub, to collect the datasets. In order to test the ability of the LLM agent, we have collected an additional dataset from Hugging Face forums. The process of collecting data from each site is illustrated below. 

\begin{table}[h]
    \centering
    \caption{Descriptions of Bug Types}
    \begin{tabular}{|p{1.8cm}|p{6.3cm}|}
    \hline
        \textbf{Bug Type} & \textbf{Description} \\ \hline
        Logic Bug (LB) & The bug indicates a lack of logical understanding in the pipeline. This bug includes using a function that does not fit the specific task, missing or incorrectly implementing any code segment, or the absence of proper guarding conditions~\cite{islam2019comprehensive}. \\ \hline
        Configuration Bug (CB) & This bug indicates any configuration errors, including parameter misconfiguration or environment misconfiguration \cite{han2016empirical}. \\ \hline
        Initialization Bug (IB) & Any variable/function may require proper initialization before use. Not initializing the variables/functions correctly before using them or initializing them wrongly (e.g., inside a loop in the wrong way) will cause this error~\cite{zhai2020ubitect}. \\ \hline
        Argument Bug (AB) & This bug indicates issues with the API signature, such as the API expecting specific arguments while the user provides arguments in an incorrect format, or passes extra or missing parameters. It is worth noting that passing an incorrect value to a function that matches its signature will not trigger this bug~\cite{rice2017detecting}. \\ \hline
        Parsing Bug (PRB) & This LLM-agent-specific bug occurs while parsing the LLM output. Often, the LLM-generated output format does not align with the parser's structure or the user's expectations, triggering the bug. It mostly occurs in the parser of the LLM agent. \\ \hline
        Prompting Bug (PPB) & This bug is related to the prompts in the LLM. This includes missing prompt variables/components or invoking the LLM with incorrect instructions. \\ \hline
        API Bug (APIB) & This bug is related to the APIs or libraries used to build the agent. Causes include dependency conflicts, incorrect version usage, bugs within the libraries themselves, or trying to use a library without installing it~\cite{islam2019comprehensive}. \\ \hline
        Reference Bug (RB) & In agentic AI, often different libraries implement modules with similar names. This bug occurs when the user refers to a different, deprecated, or missing module from a specific library. This bug mostly occurs in the import stage. \\ \hline
        Availability Bug (AVB) & This bug usually comes from the API or server side, indicating that a particular language model or service is unavailable—either because of a server issue or because the feature has not been released yet. \\ \hline
        Model Bug (MB) & This bug is related to the LLM itself and occurs when a user requests a task the LLM cannot perform, such as asking a chat model to generate an image or using a non-functional model to produce a function call.\\ \hline
        Resource Limitation Bug (RLB) & This bug is related to the user’s local system and its resource usage. It may occur when attempting to run a large LLM with insufficient system resources or limited credits. \\ \hline
    \end{tabular}
    \label{tab:bug-type-defination}
\end{table}

\begin{table}[h]
\centering
\footnotesize
\caption{Descriptions of Root Causes and Their Subclasses}
\resizebox{\columnwidth}{!}{
\begin{tabular}{|m{.8cm}|>{\raggedright\arraybackslash}m{1.5cm}|p{6.3cm}|}
\hline
\textbf{Root-Cause} & \textbf{Subclass} & \textbf{Definition} \\
\hline
        
         \multirow{2}{=}{AM} 
        & Wrong API Context & The API is used in an inappropriate situation, or there is a misunderstanding of its intended purpose. For example, using a translation API to extract sentiment~\cite{ferles2021verifying}.\\ \cline{2-3}
        & Invalid API Arguments & Incorrect types or number of arguments are passed to the API. For instance, passing a string where a JSON object is expected~\cite{humbatova2020taxonomy}. \\ \hline
        
         \multirow{2}{=}{\shortstack[l]{IMP}}
        & Incorrect Value & A valid parameter is passed, but its value is logically wrong, like using a learning rate of 10 instead of 0.01~\cite{pan2024lost}.\\ \cline{2-3}
        & Missing Value & One or more optional parameters are omitted entirely, often leading to unexpected default behavior~\cite{pan2024lost}. \\ \hline
        
         \multirow{2}{=}{\shortstack[l]{IDF}} 
        & Input Data Format Error & An error that occurs when the data from an external resource provided to the LLM is not in the expected type, structure, or schema. For example, the input text contains a stop sequence that causes the LLM to terminate its response early.  \\ \cline{2-3}
        & Output Data Format Error & The expected output format from the LLM is not met. For example, when prompted for structured output, the LLM-generated response may miss required keys or deviate from the format, causing the parser to fail when processing it. \\ \hline
        
         \multirow{2}{=}{\shortstack[l]{IMCF}} 
        & Missing Flow & A required logical code segment is not implemented, such as a missing if condition or function call~\cite{jiang2007context}. \\ \cline{2-3}
        & Incorrect Flow & This occurs when the logic is present in the code but implemented incorrectly, such as using an if condition that checks the wrong variable~\cite{islam2019comprehensive}. \\ \hline
        
         \multirow{2}{=}{\shortstack[l]{II}} 
        & Prompt Specification & Errors in constructing the prompt, such as ambiguous language, poor formatting, or missing instruction cues. It includes four issues in the prompt, including missing context, missing specifications, multiple contexts, and unclear instructions \cite{ehsani2025towards}. \\ \cline{2-3}
        & Prompt Orchestration & Logical flaws in the prompt, such as not passing variables correctly or providing a plain string when a JSON-formatted input is required.\\ \hline
        
        \multicolumn{2}{|m{2.1cm}|}{AL} 
        & Task failure due to limitations of the API or Library, such as unsupported capabilities or service downtime. Users typically cannot control this~\cite{9438601}. \\ \hline

        \multicolumn{2}{|m{2.1cm}|}{CM} 
        & Incorrect usage or selection of components like LLM, tools, or memory modules, resulting in integration failure~\cite{wang2023compatibility}. \\ \hline

        \multicolumn{2}{|m{2.1cm}|}{RV} 
        & Dependency conflicts or missing requirements, such as incompatible library versions or unmet prerequisites~\cite{catolino2019not}. \\ \hline

        \multicolumn{2}{|m{2.1cm}|}{Others} 
        & Issues not related to LLM agents, often caused by the user’s machine, environment, or external components. \\ \hline

    \end{tabular}
}
\label{tab:root-cause-definitions}
\end{table}
\renewcommand{\arraystretch}{1.0}

\begin{table}[h]
    \centering
 \footnotesize
    \caption{Descriptions of Effects}
    \begin{tabular}{|p{1.8cm}|p{6.2cm}|}
    \hline
        \textbf{Effects} & \textbf{Description} \\ \hline
        Crash  & The program throws an error and stops working~\cite{islam2019comprehensive}. \\ \hline
        Incorrect Output (IO) & LLM generates a complete output, but it does not align with the expected output~\cite{islam2019comprehensive}. \\ \hline
        Empty Response (ER) & The system does not generate an output at all~\cite{islam2019comprehensive}. \\ \hline
        Output Dump (OD) & This effect indicates that the system generated the entire output at once instead of streaming it gradually. \\ \hline
        Stateless Interaction (SI) & The agent does not remember the past conversations and answers only the current question without understanding the full context. \\ \hline
        Partial Output (PO) & The LLM generates incomplete or truncated output~\cite{islam2019comprehensive}. \\ \hline
        Tool Ignored (TI) & The effect indicates that the system does not invoke tool(s) during the process~\cite{winston2025taxonomy}. \\ \hline
        Slow Output (SO) & The output is generated slowly~\cite{tan2014bug}. \\ \hline
        Warning  & This means the program does not terminate but instead issues a warning~\cite{tan2014bug}. \\ \hline
        Hang & The system ceases to respond and requires human intervention to restore functionality~\cite{islam2019comprehensive}. \\ \hline
        Indeterminate Loop (IL) & The system runs into a loop indefinitely~\cite{islam2019comprehensive}. \\ \hline
        Resource Overuse (RO) & This effect indicates that the system consumed high resources like RAM~\cite{tan2014bug}. \\ \hline
        Silent Fail (SF) & This effect illustrates that the agent has failed to perform a task, but did not provide any log that the task failed~\cite{lou2022demystifying}. \\ \hline
        Unknown & The post does not mention the effect of the problem~\cite{tan2014bug}. \\ \hline
    \end{tabular}
    \label{tab:effect_desc}
\end{table}
\subsubsection{Stack Overflow}

 We selected Stack Overflow, one of the most popular Q\&A sites for developers \cite{moutidis2021community}, to collect our dataset. As illustrated in Figure \ref{fig:workflow}, the data collection process consists of three main stages. In the first stage, keyword searching, we gathered posts related to LLM agents by searching with keywords such as \texttt{LLM agent}, \texttt{LangChain}, \texttt{LlamaIndex}, \texttt{Semantic Kernel}, \texttt{CrewAI}, \texttt{AutoGen}, and \texttt{OpenAI}, collecting a total of 2929 posts. While each keyword represents a separate framework, \texttt{OpenAI} refers to custom agent implementations.
 
 In the second stage, we manually evaluated each post, eliminating those that were irrelevant or consisted only of questions, and posts with insufficient or no code, and removed duplicate entries. We retained only those posts that contained sufficient information about the problem. Finally, we selected 951 posts relevant to our study. In the third step, two PhD students independently analyzed each post and assigned labels. The details of data annotation are mentioned in Section \ref{sec:data-annotation}.

\subsubsection{Github}
We have collected data from two GitHub sources, namely GitHub commits and GitHub Issues, to verify the existence of bugs that we identified in the Stack Overflow dataset. For both GitHub commits and issues, we collected 1891 GitHub repositories utilizing the following libraries: LangChain, LangChain-js, LangGraph, LlamaIndex, CrewAI, AutoGen, and Semantic Kernel. These libraries were selected based on their popularity, as measured by GitHub stars \cite{awesome-llm-agents}. To identify relevant repositories, we searched GitHub Topics using each library name as the topic keyword, and scraped the resulting repositories using Selenium. We then excluded repositories maintained by the original developers of these libraries to ensure our analysis focused exclusively on bugs in the implementation of agents, rather than bugs within the libraries themselves. While the scraped repositories may include sample projects and course materials, such repositories typically do not contain closed issues or fix commits, and therefore our human annotators confirmed that the selected commits and issues in our dataset do not include repositories of that type. The subsequent steps differ between GitHub commits and issues.

\textit{GitHub Commits:}
\label{sec:git-commits-data-collection}
After collecting the repositories, we filtered for commits explicitly marked as fix commits (i.e., containing the keyword ``fix"). Following prior work \cite{islam2019comprehensive}, we randomly selected 100 commits per library and ensured mutual exclusivity across repositories that share multiple frameworks. Each commit was independently labeled by two annotators using the same bug type, root cause, and effect definitions established for the Stack Overflow dataset, with agreement evaluated at multiple stages. In total, we collected 36, 16, 16, 19, 12, 20, and 23 bugs from LangChain, LangChain-js, LangGraph, LlamaIndex, CrewAI, AutoGen, and Semantic Kernel, respectively.

\textit{GitHub Issues:}
\label{sec:git-issues-data-collection}
To collect GitHub issues, we have iterated through the repositories and collected all the closed issues. After that, similar to GitHub commits, we randomly chose 100 non-overlapping issues, if a library had more than 100 issues. Subsequently, we filtered out the relevant instances and labeled them following the classification steps defined previously.

\subsubsection{Hugging Face Forums}
Since Hugging Face plays a critical role in building LLM agents, especially with open-source LLMs \cite{hussain2024tutorial}, the Hugging Face forums contain issues related to LLM agents. Therefore, we have selected posts from the forums by searching the keywords: LangChain, LangChain-js, LangGraph, LlamaIndex, CrewAI, AutoGen, and Semantic Kernel. Thereafter, we removed the duplicate posts, and the filtered relevant posts were selected for labeling. The dataset is used for evaluation purposes only. Unlike Stack Overflow and GitHub, which were labeled by two human annotators, this dataset was labeled by one annotator. A second annotator then acted as a reviewer. The reviewer examined both the first annotator's labels and the agent's predicted labels, and independently assigned their own final labels. These final labels could agree with either the human annotator or the agent, or disagree with both.

\subsubsection{Data Annotation}
\label{sec:data-annotation}
Since no predefined taxonomy existed for annotating the dataset \cite{nistor2024haskell}, we adopted methodologies from prior works \cite{humbatova2020taxonomy, ning2024defining}, in which each author independently assigned a card and description to a given post, with final labels defined through open card sorting \cite{zampetti2021self}. The resulting bug types, root causes, and effects are presented in Tables \ref{tab:bug-type-defination}, \ref{tab:root-cause-definitions}, and \ref{tab:effect_desc}, with associated examples available in \cite{AnonymousRepo12LLMAgentBugStudy}. Several labels coincide with categories from prior literature, as discussed in Section \ref{sec:new-bugs} alongside their distributions in agentic and non-agentic systems. Beyond bug types, root causes, and effects, we report the programming languages and libraries used to build agents, as well as the agent component that initiated each bug, based on four components from prior literature \cite{ning2024defining}. Two additional labels are included in the component section, where \textit{Other} indicates the bug originated from a component external to the LLM agent, and \textit{Unknown} indicates the annotators could not determine the responsible component. To understand bug origins and causes, we followed a consistent mechanism across sources. For GitHub commits, the pre-commit version represents the buggy code and the post-commit version reflects the fix. For issues, the fix is documented via a pull request or repository owner explanation. For Stack Overflow and HuggingFace posts, verified answers were used, where a verified answer includes an accepted answer, a solution from the original user, a user-confirmed solution in comments, or an annotator-verified solution. When no verified solution was found, external sources were consulted and referenced in the dataset. Each entry also includes a rationale for the selected labels.

 Once all the labels are defined using the open card sorting method~\cite{zampetti2021self}, following a previous study \cite{islam2019comprehensive}, two PhD students independently annotated all four labels (i.e., bug type, root cause, effect, and component). For Stack Overflow, the programming language and framework were identified using tags, while for GitHub, data was collected based on the frameworks. Therefore, these two labels were annotated collaboratively, in contrast to the other four labels, which were annotated independently. Agreement between the two annotators was measured at 5\%, 10\%, 20\%, 30\%, 40\%, 50\%, 60\%, 70\%, 80\%, 90\%, and 100\% intervals using Cohen's Kappa coefficient \cite{landis1977measurement}, with disagreements addressed after each iteration. When the two annotators failed to reach an agreement, a moderator with expertise in the field adjudicated. At the last iteration, the agreement reached 0.93, 0.95, 0.96, and 1.00 for bug type, root cause, effect, and component, respectively. { The agreement scores at different iteration stages for Stack Overflow posts are reported in \cite{AnonymousRepo12LLMAgentBugStudy}.}

At the initial labeling stage, there was slight agreement between the annotators, which increased to a moderate level across all components by 30\% of the data. In subsequent iterations, the annotators achieved perfect agreement across all labels for the Stack Overflow dataset. Due to the limited number of samples, agreement for GitHub issues and commits was measured after annotations were completed. Since the GitHub data was annotated after the Stack Overflow data, the level of disagreement was low. For GitHub issues, the kappa scores for bug type, root cause, and component were 0.9422, 0.7084, and 0.983, respectively, while the effect showed no disagreement. For GitHub commits, no disagreement was recorded. Since some posts may contain multiple bugs, instances with multiple bug types or root causes were annotated multiple times, {resulting in 15 such instances and 1,283 total annotations}. Lastly, since the Hugging Face forum data was used only for evaluation, one annotator labeled it, agreement was measured against the best LLM, and a second reviewer later checked both.

\subsection{Most common bug type, root cause and effect (RQ1)}
While bug distributions vary across platforms, the majority of bug types, root causes, and effects are common to all. Logic Bug is the primary bug type in both Stack Overflow and GitHub, accounting for 22.6\% and 30.2\% of all bugs, respectively. Table \ref{tab:data-distribution} presents the distribution of bug types, root causes, and effects.

\definecolor{LB}{RGB}{100,180,200}     
\definecolor{CB}{RGB}{80,100,170}      
\definecolor{APIB}{RGB}{80,160,80}     
\definecolor{IB}{RGB}{60,80,150}       
\definecolor{AB}{RGB}{140,190,140}     
\definecolor{RB}{RGB}{160,140,200}     
\definecolor{PPB}{RGB}{100,60,140}     
\definecolor{RLB}{RGB}{120,170,190}    
\definecolor{MB}{RGB}{170,200,170}     
\definecolor{AVB}{RGB}{190,210,190}    
\definecolor{PRB}{RGB}{200,210,230}    

\definecolor{CR}{RGB}{100,130,190}     
\definecolor{IO}{RGB}{130,100,180}     
\definecolor{UN}{RGB}{100,170,100}     
\definecolor{ER}{RGB}{80,150,80}       
\definecolor{TI}{RGB}{170,200,170}     
\definecolor{OD}{RGB}{190,190,190}     
\definecolor{SI}{RGB}{160,180,210}     
\definecolor{SF}{RGB}{200,185,210}     
\definecolor{PO}{RGB}{180,200,180}     
\definecolor{SO}{RGB}{210,195,215}     
\definecolor{WR}{RGB}{210,210,225}     
\definecolor{IL}{RGB}{220,215,230}     
\definecolor{RO}{RGB}{150,120,170}     
\definecolor{HG}{RGB}{200,210,220}

\definecolor{LB}{RGB}{180,210,220}
\definecolor{CBc}{RGB}{170,180,210}
\definecolor{APIB}{RGB}{170,200,170}
\definecolor{IBc}{RGB}{160,175,205}
\definecolor{AB}{RGB}{195,215,195}
\definecolor{RB}{RGB}{210,200,225}
\definecolor{PPB}{RGB}{185,170,205}
\definecolor{RLB}{RGB}{185,205,215}
\definecolor{MB}{RGB}{205,220,205}
\definecolor{AVB}{RGB}{215,225,215}
\definecolor{PRB}{RGB}{215,220,230}
\definecolor{CRc}{RGB}{180,195,220}
\definecolor{IO}{RGB}{195,180,215}
\definecolor{UN}{RGB}{180,210,180}
\definecolor{ERc}{RGB}{170,205,170}
\definecolor{TI}{RGB}{205,220,205}
\definecolor{OD}{RGB}{210,210,210}
\definecolor{SI}{RGB}{200,210,220}
\definecolor{SF}{RGB}{215,205,220}
\definecolor{PO}{RGB}{205,215,205}
\definecolor{SOc}{RGB}{220,210,222}
\definecolor{WR}{RGB}{220,220,230}
\definecolor{IL}{RGB}{225,220,230}
\definecolor{RO}{RGB}{195,180,210}
\definecolor{HG}{RGB}{210,215,222}
\definecolor{IMCFIF}{RGB}{180,210,215}
\definecolor{RVc}{RGB}{190,200,220}
\definecolor{AL}{RGB}{180,205,180}
\definecolor{AMIAA}{RGB}{160,175,205}
\definecolor{IMCFMF}{RGB}{195,215,210}
\definecolor{IMPIV}{RGB}{200,215,200}
\definecolor{IMPMV}{RGB}{205,218,205}
\definecolor{Others}{RGB}{200,200,200}
\definecolor{CM}{RGB}{205,205,218}
\definecolor{AMWAC}{RGB}{195,185,210}
\definecolor{IIPS}{RGB}{185,175,205}
\definecolor{IIPO}{RGB}{195,183,208}
\definecolor{IDFID}{RGB}{215,208,220}
\definecolor{IDFOD}{RGB}{220,215,225}

\begin{table*}[]
\caption{Distribution of bug types, root causes, and effects across Stack Overflow and GitHub.}
\label{tab:data-distribution}
\scriptsize
\setlength{\tabcolsep}{2pt}
\newcolumntype{W}{>{\raggedright\arraybackslash}p{0.8cm}}
\newlength{\dcolw}
\setlength{\dcolw}{\dimexpr(\textwidth-.55cm-0.8cm-32\tabcolsep-17\arrayrulewidth)/14\relax}
\newcolumntype{Z}{>{\raggedleft\arraybackslash}p{\dcolw}}
\newcolumntype{Y}{>{\centering\arraybackslash}p{\dcolw}}
\begin{tabular}{|p{.55cm}|W|*{14}{Z|}}
\hline
\multirow{3}{.55cm}{\centering Bug Type} &
Label &
\multicolumn{1}{Y|}{\cellcolor[HTML]{B4D2DC}LB} &
\multicolumn{1}{Y|}{\cellcolor[HTML]{AAC8AA}APIB} &
\multicolumn{1}{Y|}{\cellcolor[HTML]{80AFCD}CB} &
\multicolumn{1}{Y|}{\cellcolor[HTML]{C3D7C3}AB} &
\multicolumn{1}{Y|}{\cellcolor[HTML]{B9AACD}PPB} &
\multicolumn{1}{Y|}{\cellcolor[HTML]{A0AFCD}MB} &
\multicolumn{1}{Y|}{\cellcolor[HTML]{B9CDD7}RB} &
\multicolumn{1}{Y|}{\cellcolor[HTML]{B4C3DC}PRB} &
\multicolumn{1}{Y|}{\cellcolor[HTML]{B4D2DC}RLB} &
\multicolumn{1}{Y|}{\cellcolor[HTML]{A0AFCD}IB} &
\multicolumn{1}{Y|}{\cellcolor[HTML]{D7E1D7}AVB} &
\cellcolor[HTML]{C0C0C0}\mbox{} & \cellcolor[HTML]{C0C0C0}\mbox{} & \cellcolor[HTML]{C0C0C0}\mbox{} \\
\cline{2-16}
& SO & 22.6 & 18.8 & 18.1 & 11.2 & 6.3 & 5.9 & 5.4 & 3.6 & 3.6 & 2.6 & 1.8 & \cellcolor[HTML]{C0C0C0}\mbox{} & \cellcolor[HTML]{C0C0C0}\mbox{} &\cellcolor[HTML]{C0C0C0}\mbox{} \\
\cline{2-16}
& GH & 30.2 & 16.7 & 16.7 & 6.0 & 5.6 & 2.3 & 6.0 & 0.5 & 4.2 & 10.7 & 0.9 & \cellcolor[HTML]{C0C0C0}\mbox{} & \cellcolor[HTML]{C0C0C0}\mbox{} & \cellcolor[HTML]{C0C0C0}\mbox{} \\
\hline\hline
\multirow{3}{.55cm}{\centering Root Cause} &
Label &
\multicolumn{1}{Y|}{\cellcolor[HTML]{B4D2D7}IMCF.IF} &
\multicolumn{1}{Y|}{\cellcolor[HTML]{C3D7D2}IMCF.MF} &
\multicolumn{1}{Y|}{\cellcolor[HTML]{C8D7C8}IMP.IV} &
\multicolumn{1}{Y|}{\cellcolor[HTML]{CDDACC}IMP.MV} &
\multicolumn{1}{Y|}{\cellcolor[HTML]{A0AFCD}AM.IAA} &
\multicolumn{1}{Y|}{\cellcolor[HTML]{C3B9D2}AM.WAC} &
\multicolumn{1}{Y|}{\cellcolor[HTML]{B9AFCD}II.PS} &
\multicolumn{1}{Y|}{\cellcolor[HTML]{C3B7D0}II.PO} &
\multicolumn{1}{Y|}{\cellcolor[HTML]{D7D0DC}IDF.ID} &
\multicolumn{1}{Y|}{\cellcolor[HTML]{DCD7E1}IDF.OD} &
\multicolumn{1}{Y|}{\cellcolor[HTML]{CDCDDA}CM} &
\multicolumn{1}{Y|}{\cellcolor[HTML]{BED2DC}RV} &
\multicolumn{1}{Y|}{\cellcolor[HTML]{B4CDB4}AL} &
\multicolumn{1}{Y|}{\cellcolor[HTML]{C8C8C8}Others} \\
\cline{2-16}
& SO & 14.4 & 8.7 & 8.6 & 8.0 & 9.7 & 4.8 & 3.5 & 2.6 & 1.3 & 0.9 & 5.4 & 14.1 & 10.8 & 7.1 \\
\cline{2-16}
& GH & 34.4 & 6.5 & 12.1 & 4.7 & 8.8 & 2.8 & 4.7 & 0.9 & 0.5 & 0.5 & 2.8 & 13.0 & 3.3 & 5.1 \\
\hline\hline
\multirow{3}{.55cm}{\centering Effect} &
Label &
\multicolumn{1}{Y|}{\cellcolor[HTML]{B4C3DC}CR} &
\multicolumn{1}{Y|}{\cellcolor[HTML]{C3B4D7}UN} &
\multicolumn{1}{Y|}{\cellcolor[HTML]{C8D2DC}IO} &
\multicolumn{1}{Y|}{\cellcolor[HTML]{D7CDDC}SF} &
\multicolumn{1}{Y|}{\cellcolor[HTML]{DCDCE6}WR} &
\multicolumn{1}{Y|}{\cellcolor[HTML]{D7E1D7}SO} &
\multicolumn{1}{Y|}{\cellcolor[HTML]{CDDCCD}TI} &
\multicolumn{1}{Y|}{\cellcolor[HTML]{E1DCE6}IL} &
\multicolumn{1}{Y|}{\cellcolor[HTML]{CDD7CD}PO} &
\multicolumn{1}{Y|}{\cellcolor[HTML]{C3B4D2}RO} &
\multicolumn{1}{Y|}{\cellcolor[HTML]{D2D7DE}SI} &
\multicolumn{1}{Y|}{\cellcolor[HTML]{D2D2D2}ER} &
\multicolumn{1}{Y|}{\cellcolor[HTML]{DCD7E1}OD} &
\multicolumn{1}{Y|}{\cellcolor[HTML]{D2D7DE}HG} \\
\cline{2-16}
& SO & 60.6 & 6.6 & 16.6 & 1.6 & 1.0 & 1.1 & 2.2 & 0.9 & 1.3 & 0.7 & 1.6 & 3.7 & 1.7 & 0.2 \\
\cline{2-16}
& GH & 47.0 & 27.0 & 14.9 & 5.6 & 1.4 & 0.9 & 0.9 & 0.5 & 0.5 & 0.5 & 0.5 & 0.5 & 0.0 & 0.0 \\
\hline
\end{tabular}
\end{table*}

Although logic bugs also occur in traditional software, existing tools like Pylint \cite{pylint} cannot be directly applied to LLM agents due to their different reasoning mechanisms and require further adaptation. While parsing and model bugs are less common in GitHub, they appear more frequently in Stack Overflow, likely because GitHub commits reflect already-built systems, whereas parsing and model bugs tend to arise during the initial implementation phase.

Unlike bug types, root causes are more diverse. The primary root causes in Stack Overflow are Incorrect or Missing Control Flow-Incorrect Flow (IMCF.IF) and Requirement Violation. IMCF.IF is common because LLM agent development is a relatively new field, and developers are often unfamiliar with managing the multiple interconnected components that agent systems require. To address this, developers should map out the agent's flow before implementation and use static analysis tools such as Pylint to catch some missing or unhandled conditions. Requirement Violation is also prevalent, as LLM agent development often involves multiple evolving frameworks, making dependency management challenging. This root cause is less common in GitHub, where library versions are often fixed in finished projects. Tools like Poetry \cite{python_poetry} can help by locking project dependencies and reducing the risk of conflicting packages.

Crash is the most dominant effect across both platforms, accounting for 60.6\% of bugs in Stack Overflow and 47\% in GitHub. Unknown is also prominent in GitHub, as the effect is often difficult to determine when not explicitly mentioned in commit descriptions. While some tools localize bugs based on crash reports, their effectiveness in LLM agents remains unexplored \cite{medeiros2024impact}. Incorrect output is another common effect across all platforms, while Hang and Resource Overuse are the least frequent.

\begin{figure}[h]
    \centering
    \includegraphics[width=1\linewidth]{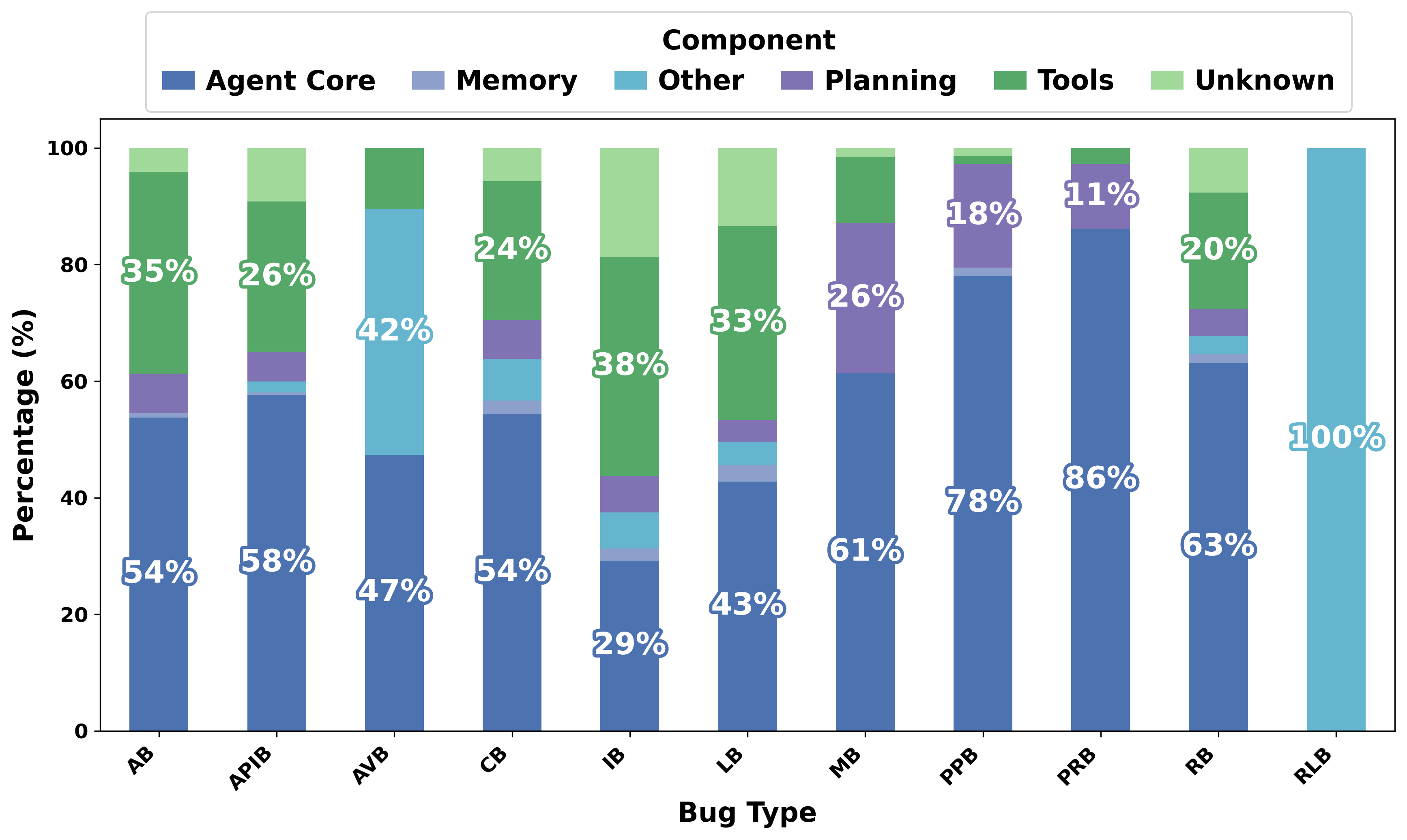}
    \caption{Component distribution across bug type.}
    \label{fig:component_per_bug_type}
\end{figure}

\vspace{-10pt}

\subsection{Most bug-prone LLM agent component (RQ2)}
\begin{figure}[h]
    \centering
    \includegraphics[width=1\linewidth]{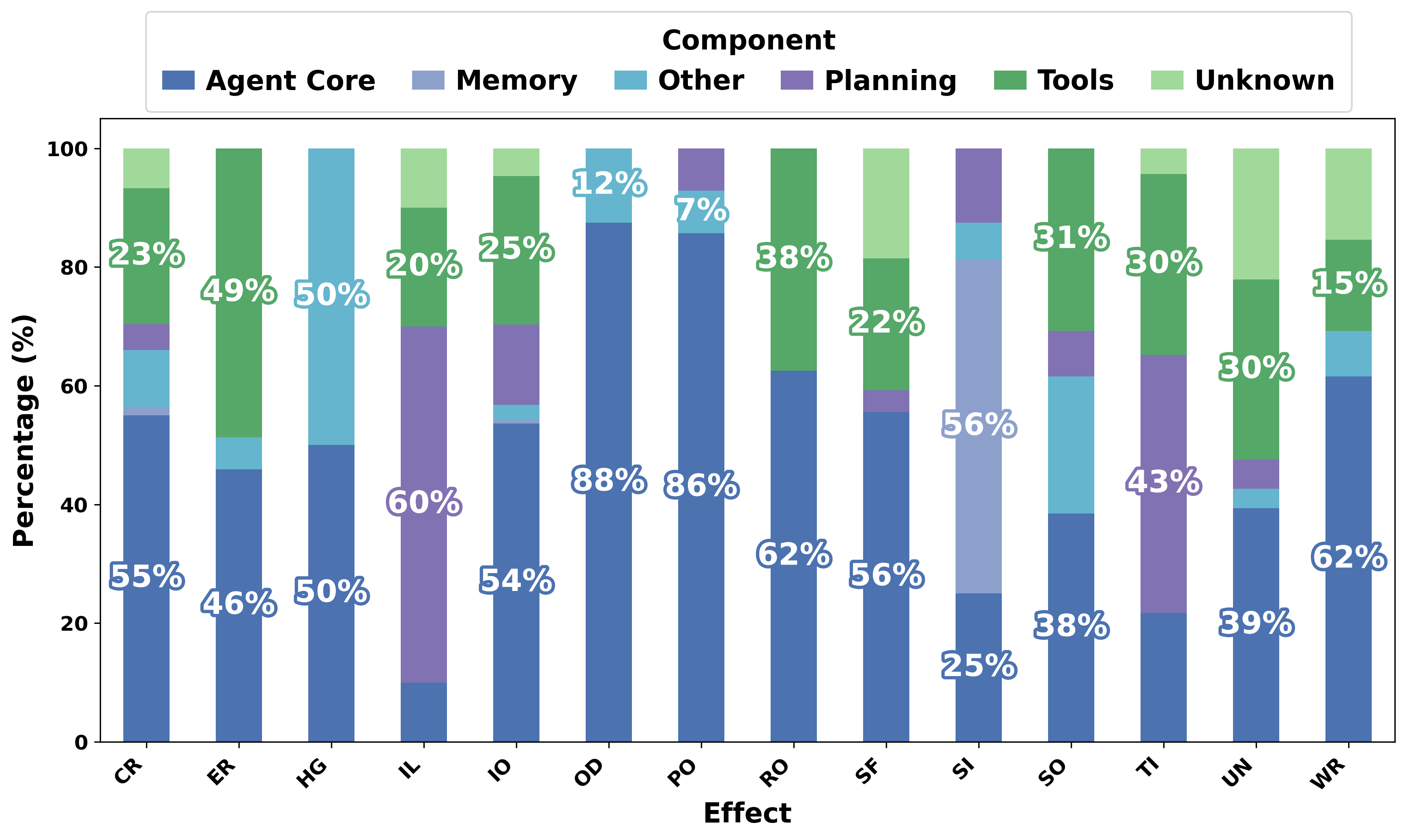}
    \caption{Component distribution across effects.}
    \label{fig:component_per_effect}
\end{figure}

Although identifying the exact component affected by a bug in GitHub is challenging due to large project sizes, the agent core is the most error-prone among known components. This aligns with Stack Overflow, where the agent core accounts for 57.1\% of bugs, followed by tools \cite{AnonymousRepo12LLMAgentBugStudy}. These results suggest that prioritizing testing, debugging, and monitoring of the agent core can significantly reduce bug likelihood. The distribution of components across bug types and effects is provided in Figures \ref{fig:component_per_bug_type}, and \ref{fig:component_per_effect}, respectively, while the distribution of root causes per component is shared in \cite{AnonymousRepo12LLMAgentBugStudy}. As shown in Figure \ref{fig:component_per_bug_type}, the agent core accounts for more than half of the bugs in every bug type, except for Initialization and Resource Limitation bugs. Notably, Resource Limitation bugs do not fall under any agent component, as they are not caused by the LLM agent itself.

 According to the distribution of root causes and effects across components, the agent core is the primary component where bugs occur, except for those caused by Incorrect Data Flow (Input Data Format Error) and Others. While identifying bug types and root causes from the output alone is difficult, determining effects is comparatively easier. Agents exhibiting indeterminate loops encounter issues in the planning component in 66.7\% of cases, whereas stateless interactions stem from bugs in the memory component in 53.33\% of cases. Therefore, observed effects can serve as useful indicators for bug localization, allowing developers to narrow their focus to the relevant components and speed up debugging.

\vspace{-14pt}

\subsection{Distribution between bugs (RQ3)}
\begin{figure}[h]
    \centering
    \includegraphics[width=1\linewidth]{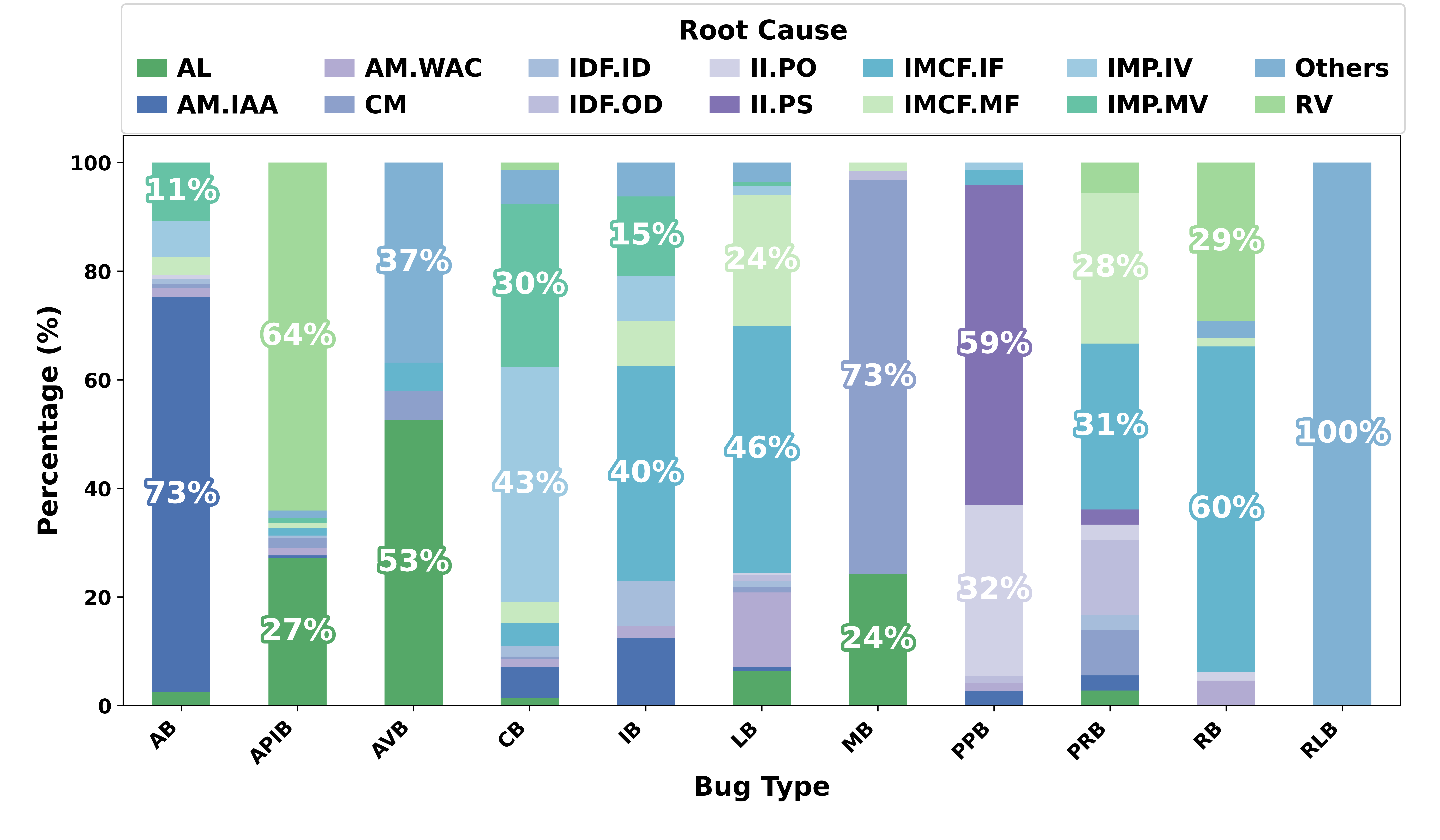}
    \caption{Distribution of root cause across bug types.}
    \label{fig:root_cause_per_bt}
\end{figure}

To answer RQ3, we analyzed the correlation between bug types and their root causes. As shown in Figure \ref{fig:root_cause_per_bt}, most API-related bugs occur due to requirement violations, driven by the rapid release of new library versions and conflicting dependencies across multiple frameworks. LangChain, the most commonly used library~\cite{awesome-llm-agents}, is particularly susceptible to such errors. Tools like Poetry \cite{python_poetry} can lock library versions, while virtual environments can resolve some conflicts. The second most common cause is API limitations, attributable to rapid library development with limited testing. Argument Bugs, the fourth most common type, predominantly occur due to API misuse, as developers are often unfamiliar with existing method signatures in this emerging field, though their proportion has been decreasing since 2023 as shown in RQ4. Most prompt-related bugs arise from incorrect prompt structure, while fewer stem from prompt logic issues such as incorrectly passing prompt variables. Lastly, resource limitation bugs are categorized under `Others' root causes, as they are not directly related to agent development.

\subsection{Bug distribution over time (RQ4)}
To evaluate the distribution of bugs across different years, we analyzed the six characteristics based on the post date. According to the analysis presented in \cite{AnonymousRepo12LLMAgentBugStudy}, during 2021 and 2022, which marked the early stage of LLM agents, only a small number of Stack Overflow posts related to bugs in LLM agents were made, totaling fewer than 30 in the collected dataset. At the end of 2022, with the release of libraries for building LLM agents, for instance, LangChain in October 2022, the development of LLM agents surged, resulting in the highest number of bug-related posts on Stack Overflow in 2023. In our dataset, 2024 and 2025 show fewer posts related to bugs in LLM agents. This may reflect fewer development challenges, a slight drop in development activity, framework changes that reduced errors, or tools like ChatGPT helping developers resolve bugs on their own. Further investigation is needed to understand this trend. The number of posts in 2026 does not reflect the entire year, as data collection stopped after March 2025.
\vspace{-10pt}

As shared in \cite{AnonymousRepo12LLMAgentBugStudy}, the proportion of reported bugs in JavaScript has been decreasing since 2023, with 14.56\% of bugs reported in 2023 coming from JavaScript, dropping to 7.30\% in 2025. In contrast, the proportion of bugs reported in C\# has been increasing steadily during the same period, rising from 2.63\% to 5.11\%. The trend may suggest that developers use JavaScript less for building agents, or that they encounter fewer bugs when doing so, though further investigation is needed to understand the trend.

Most major libraries, such as LangChain, Autogen, and CrewAI, were released at the end of 2022, after which developers began adopting them over custom-made agents. As shared in \cite{AnonymousRepo12LLMAgentBugStudy}, LangChain remains the most popular framework. Semantic Kernel usage grew from 0.95\% in 2023 to 5.84\% in 2025, and LangGraph gained significant traction as multi-agent systems became more complex, with its share rising from 1.64\% in 2024 to 5.84\% in 2025. This trend highlights the need for automatic code generation and optimization tools for LLM agents, similar to Auto-Keras for DNNs \cite{jin2019auto}.

Since LangChain is the most popular framework for building LLM agents~\cite{awesome-llm-agents}, we investigated the common root causes of bugs in LangChain-based agents. As detailed in \cite{AnonymousRepo12LLMAgentBugStudy}, API limitations and API misuse accounted for 25.22\% of bugs in 2023, rising to 29.86\% in 2025. A similar trend is observed in CrewAI \cite{AnonymousRepo12LLMAgentBugStudy}, where this proportion increased from 22.23\% in 2024 to 60\% in 2025. An increasing number of API limitation bugs is also reported in LangChain-js, LlamaIndex, Semantic Kernel, and Autogen, driven by the rapid evolution of these libraries. Library maintainers should prioritize rigorous documentation before releasing new versions. To address API misuse, researchers should treat agentic libraries and general-purpose libraries as distinct categories, similar to efforts in the deep learning field \cite{yang2024demystifying}.

\vspace{-1pt}
\subsection{Cross-System Bugs Comparison}
\label{sec:new-bugs}
Our study identifies several bug types, root causes, and effects that also appear in other software systems, such as Logic and Initialization Bugs, API Misuse, and Crash. However, we also identify categories unique to agentic systems, including Parsing, Reference, Availability, Model, and Resource Limitation Bugs, as well as novel root causes such as Prompt Orchestration. These are rare in traditional software due to the distinct characteristics of agentic systems. For instance, LLMs produce dynamic output that demands flexible parsers, and frequent library reorganization causes outdated references and import errors. Availability bugs arise because failures in external LLMs or tools can bring down the entire agent. Model bugs are amplified by the diversity of LLM variants that makes it difficult to select the appropriate model for a task. Resource limitation bugs are more prevalent due to the specialized computational demands of LLMs.

Although some bug categories overlap with those in other software fields, their distribution differs significantly in agentic systems. In deep learning compiler systems, 9.78\% of bugs stem from API misuse~\cite{shen2021comprehensive}, and a study on open-source C programs reports 17.05\%, compared to 14.5\% and 11.6\% in agentic systems on Stack Overflow and GitHub, respectively. Similarly, data format bugs account for 25\% of bugs in deep learning systems~\cite{shah2025towards} but only 2.2\% in agentic systems, as agents rely on pre-trained models rather than fine-tuning or training from scratch, which reduces dependencies on input datasets and data formatting. Some similarities in bug distribution are also observed; for instance, API bugs in PyTorch-based deep learning systems account for 16\%~\cite{islam2019comprehensive}, which is consistent with the distribution observed in this study.

\subsection{Actionable Insights}
This section discusses the implications from our study (RQ1–RQ4) and provides recommendations for developers, library maintainers, and researchers to minimize bugs in LLM agents. A summary of the insights is shared in \cite{AnonymousRepo12LLMAgentBugStudy}.

\subsubsection{Developers} From RQ3, requirement violation is the most common cause of API bugs, affecting 61.88\% of this type. Since new library releases often introduce conflicting requirements in this rapidly evolving field, to address some of these issues, developers can leverage dependency and packaging tools such as Poetry \cite{python_poetry}, which defines a project's dependencies in a single file and ensures consistent package versions. Additionally, argument-related bugs, largely caused by misaligned parameters, have been decreasing since 2023, though developers can further address this by integrating type-checking tools to ensure correct method invocations \cite{oh2024towards}. Lastly, API limitation bugs account for 10.8\% of Stack Overflow bugs and have been increasing over the past three years across most libraries (RQ4). While identifying such bugs can be challenging for small teams, reviewing a library's open issues before adoption is a practical approach to gain awareness of potential challenges.

\subsubsection{Library Maintainers} As LLM agent libraries grow rapidly, new versions often conflict with existing dependencies. For instance, LlamaIndex released over 140 versions in 2024 \cite{llamaindex-pypi}. Therefore, when releasing a new version, library maintainers should not only document newly added features but also specify any underlying version changes, so developers can anticipate potential conflicts during installation. To further mitigate the increasing number of API limitation bugs highlighted in RQ4, library maintainers should conduct more extensive testing before new releases.

\subsubsection{Researchers}
As API limitations are common beyond LLM agents, multiple Software Composition Analysis (SCA) tools exist to report known vulnerabilities, licensing issues, and outdated components \cite{sharma2024understanding}. OSV-Scanner \cite{osv_scanner}, developed by Google, is one popular tool in this domain. It queries the OSV database \cite{osv} to identify vulnerabilities. However, while the database includes some known vulnerabilities of LLM agent libraries such as LangChain, OSV-Scanner often fails to capture risks specific to these libraries. To the best of our knowledge, existing SCA tools are not well-suited for LLM agent libraries. Researchers could extend existing SCA tools or develop new ones that aggregate known issues from community platforms such as Stack Overflow, Hugging Face forums, OpenAI Discussions, or Reddit, to provide developers with a more comprehensive understanding of potential vulnerabilities before adopting a library.

\vspace{-5pt}

\section{LLM Agent for Labeling Bugs}
\label{sec:llm-agent}
\begin{figure*}[h]
    \centering
    \includegraphics[height=5cm,width=1\linewidth]{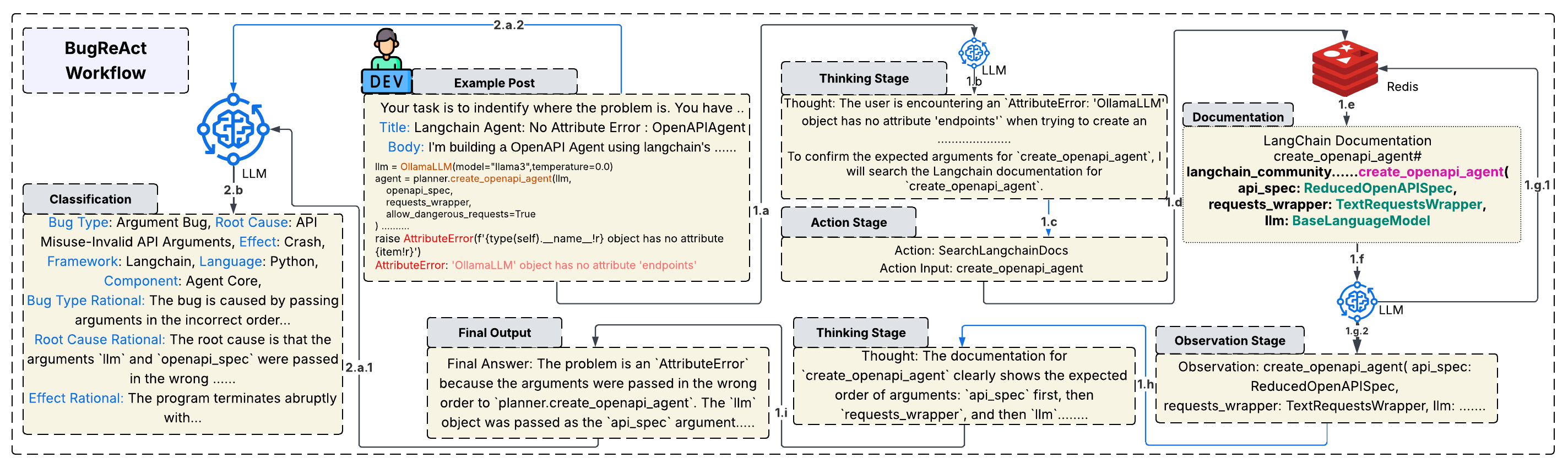}
    \caption{The flow of automatic annotation for a Stack Overflow post \cite{badhusha2025langchain} through \name, equipped with Gemini 2.5 Flash.}
    \label{fig:example}
\end{figure*}
This section describes the LLM agent designed for annotating bugs. As shown in Figure \ref{fig:workflow}, the agent consists of two main components. The first component is a~\react~agent that leverages different tools to reason about where the bug occurred. The second component takes this reasoning and maps it to the appropriate class labels. The subsequent sections explain the two components of the agent in detail, along with an example explaining the classification process. 
\vspace{-5pt}
\subsection{BugReAct Agent}
To annotate bugs in LLM agents, we built a~\name~agent \cite{yao2023react}. Although multiple agent types exist, including Reflexion \cite{NEURIPS2023_1b44b878}, function-calling \cite{Granite-functio-calling}, and ExpeL agents \cite{deng-etal-2025-agentpro}, \react~agents stand out for their explicit thought-action-observation loop, which simplifies tool integration and debugging \cite{Zhao_Huang_Xu_Lin_Liu_Huang_2024}. The proposed agent is equipped with 10 tools that scrape documentation for eight libraries (LangChain, LangChain-js, LangGraph, Pydantic, CrewAI, LlamaIndex, Semantic Kernel, and Autogen) and two community forums (the OpenAI developer community and GitHub Discussions, the latter covering all eight frameworks). Since documentation and forum searches can yield irrelevant results that overflow the context window, a summarization module is introduced after each search to condense retrieved content. To further improve efficiency, an in-memory database (Redis \cite{RedisPlatform}) is integrated to avoid reprocessing previously searched content, reducing retrieval delay and improving agent speed. The agent can invoke any tool multiple times, with each output treated as an observation that guides subsequent actions, until enough information is gathered to produce a final output with both a bug explanation and the reasoning behind its occurrence.
\vspace{-5pt}
\subsection{Classification}
This component takes the explanation generated by the~\name~agent along with the post or code snippet and passes them through an LLM for classification. The LLM is provided with class labels and definitions via the prompt and assigns one label per category, along with rationales for bug type, root cause, and effect. Pydantic enforces structured output, and for LLMs that do not natively support this, strict prompting is applied to ensure the required format.

\subsection{Workflow of~\name~with an Illustrative Example}
Figure \ref{fig:example} illustrates the decision-making workflow of a Stack Overflow post \cite{badhusha2025langchain} processed by \name. The process begins when the post is wrapped in a prompt and passed to the LLM agent (1.a), which analyzes the input and formulates an action plan (1.b). The agent then transitions to the action stage, where it determines that consulting the LangChain documentation is needed (1.c). The documentation tool is invoked with a specific query and first checks whether the keyword exists in the database (1.d). If not, it queries the LangChain documentation directly (1.e). Since retrieved content may be irrelevant or exceed the context window, it is passed through a summarization step (1.f). The summarized output is stored in Redis for future use (1.g.1) and returned to the agent (1.g.2). The agent interprets this as an observation and proceeds to the thought stage (1.h). If the output is insufficient, the tool is re-invoked to gather more information (1.i). Once adequate evidence is collected, the agent produces a final decision with an explanation of the bug's underlying cause, concluding the first stage of \name.

The second stage begins when the final output from Stage 1 (2.a.1) and the original input (2.a.2) are passed to the LLM, along with a predefined set of labels and their definitions. Based on these inputs, the model selects six labels describing the bug's characteristics, accompanied by rationales for bug type, root cause, and effect (2.b). The predicted labels are then compared against the human-annotated benchmark to evaluate model performance.

\vspace{-10pt}
\section{Evaluation}
\label{sec:evaluation}
We implemented~\name~in Python, using LangChain as the primary agent framework, alongside Pydantic for structured output, Selenium for web scraping, and Redis for in-memory storage. Experiments were run on a MacBook M3 with 16 GB RAM, with LLMs invoked via the OpenAI, OpenRouter, and Claude APIs. The codebase follows a modular structure with proper exception handling. We evaluated~\name~based on the following research questions:
\begin{itemize}
  
  \item \textbf{RQ5 (Effectiveness)} How effective is~\name~in characterizing bugs in LLM agents?
  \item \textbf{RQ6 (Reliability)} How accurate is~\name~in labeling LLM bugs compared to human annotators?
  \item \textbf{RQ7 (Efficiency \& Cost)} How efficient and cost-effective is~\name~in characterizing bug types, root causes, and effects?
\end{itemize}

\vspace{-5pt}
\subsection{Effectiveness}

\begin{figure}[h]
    \centering
    \includegraphics[ width=1\linewidth]{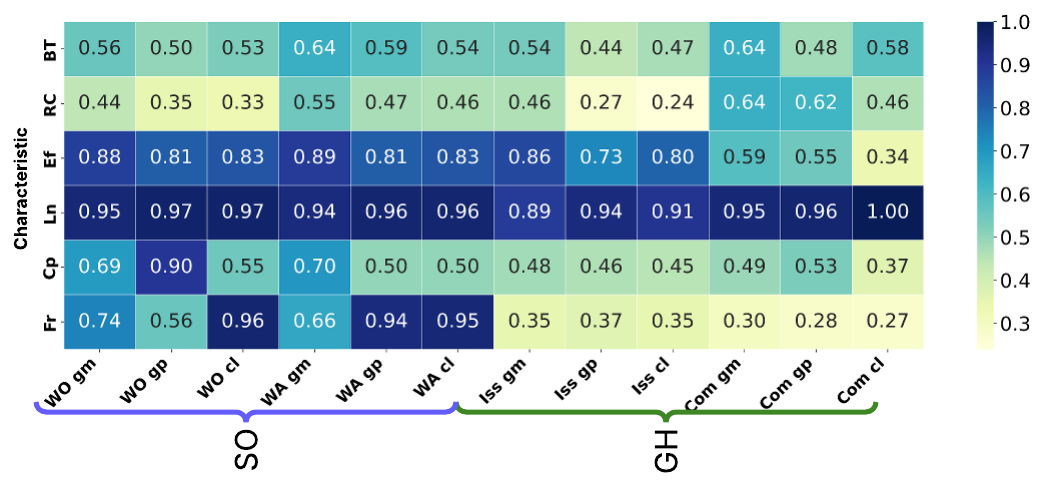}
    \caption{F1-score of different models in different datasets.}
    \label{fig:f1-score}
\end{figure}

To evaluate \name, we tested it on an annotated Stack Overflow and GitHub dataset using three powerful LLMs, namely Gemini 2.5 Flash \cite{comanici2025gemini}, GPT o3 mini \cite{openai2025o3}, and Claude Sonnet 4 \cite{anthropic2025claudesonnet4}. These models were selected for their state-of-the-art reasoning capabilities, compatibility with~\name~agents, and diversity across providers. Since the collected dataset is highly imbalanced, accuracy alone can yield a biased evaluation. Therefore, we report the F1-score, which combines precision and recall for a more balanced assessment. For a deeper analysis of the Stack Overflow dataset, we also evaluate~\name~with and without the accepted answer. Figure \ref{fig:f1-score} presents the F1-scores of the three LLMs on both datasets. Here BT: Bug Type, RC: Root Cause, Ef: Effect, Ln: Language, Cp: Component, Fr: Framework, SO: Stack Overflow, GH: GitHub, WO: without accepted answer, WA: with accepted answer, Iss: GitHub issues, Com: GitHub commits, gm: Gemini, gp: GPT, cl: Claude. 
\vspace{-7pt}

\vspace{-7pt}

\vspace{-5pt}
Gemini 2.5 Flash consistently achieves the best performance across both datasets, followed by GPT o3 mini and then Claude Sonnet 4. GPT o3 mini tends to produce shorter, more compact outputs \cite{ballon2025relationship}, but often skips necessary tool calls to verify its knowledge, unlike Gemini 2.5 Flash, which actively invokes tools to confirm specific details. Claude Sonnet 4, despite being the latest in its family, frequently over-invokes tools and struggles to converge on a correct conclusion. Our pilot study with Claude Opus 4 revealed a 4.36\% lower F1-score compared to Claude Sonnet 4.
\vspace{-12pt}
\begin{figure}[h]
    \centering
    \includegraphics[height=3.5cm,width=1\linewidth]{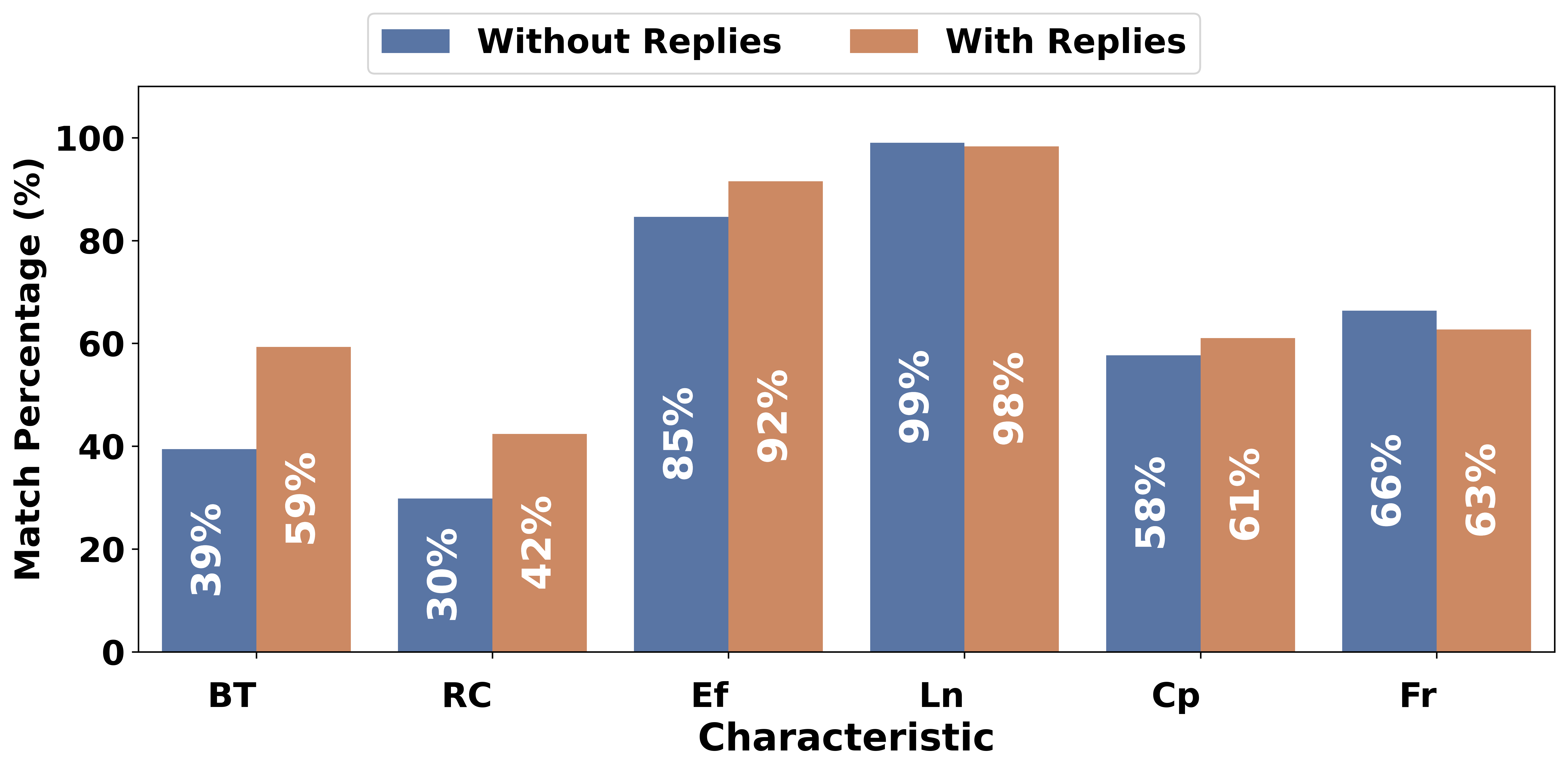}
    \caption{Comparison of \name's match with human annotators, with and without replies.}
    \label{fig:hf-comparison}
\end{figure}

\vspace{-17pt}

\subsection{Reliability}
\vspace{-5pt}
For reliability evaluation, we compare~\name~predictions with those of a human annotator on the Hugging Face Forums dataset. Figure \ref{fig:hf-comparison} shows this comparison for Gemini 2.5 Flash and GPT o3 mini.~\name~matches only 39.4\% of human annotations when provided with just the title and body of a post. This match rises to 59.3\% when replies, specifically those marked by the human annotator as solutions, are included. Agreement is higher in critical components such as root cause, effect, and LLM agent component, while less critical aspects such as language and framework show slight decreases when the accepted answer is provided. Although performance dropped for GPT o3 mini and Claude Sonnet 4,~\name~demonstrates strong agreement with the human annotator, particularly when given the full post context with replies. Whether the agent can consistently provide annotations with clear rationales, comparable to those of human annotators, remains an open question for future research. To understand our solution's failure points, we manually analyzed the instances where~\name~failed to correctly annotate bugs. In these cases, the tool is invoked with incorrect input, such as wrong search terms, which leads to a wrong assessment of the bug and, in turn, wrong labels.

To further evaluate effectiveness, a second human annotator reviewed both the human-generated and \name-generated labels and produced an independent label. The agreement table shared in \cite{AnonymousRepo12LLMAgentBugStudy} presents the agreement rates between the reviewer, the original annotator, and \name. Since Gemini 2.5 Flash achieved the best performance, only this model was used. The reviewer agreed with the human annotator in all cases except bug type (93.33\%) and root cause (97.14\%). On HuggingFace forum posts,~\name~achieved notable performance in effect annotation but lower performance in root cause annotation.

\vspace{-7pt}
\begin{table}[h]
\caption{Effectiveness (seconds) and cost (\$) of different LLMs across different platforms.}
\label{tab:effectiveness_cost}
\begin{tabular}{|
p{1.2cm}
p{1.1cm}
p{1cm}
|
>{\raggedleft\arraybackslash}p{.9cm}|
>{\raggedleft\arraybackslash}p{.9cm}|
>{\raggedleft\arraybackslash}p{.9cm}
|}
\hline
\multicolumn{2}{|c|}{\textbf{Dataset and Setting}} & \textbf{Metric}                                                                                               & \textbf{Gemini} & \textbf{GPT}   & \textbf{Claude} \\ \hline
\multicolumn{1}{|p{1.2cm}|}{\multirow{4}{*}{\parbox{1.2cm}{\raggedright Stack Overflow}}} & \multicolumn{1}{p{1.1cm}|}{\multirow{2}{*}{\parbox{1.1cm}{\raggedright Without Answer}}} & Time  & 34.00  & 19.09 & 125.93 \\ \cline{3-6} 
\multicolumn{1}{|p{1.2cm}|}{}                                & \multicolumn{1}{p{1.1cm}|}{}                                & Cost  & 0.016  & 0.017 & 0.049  \\ \cline{2-6} 
\multicolumn{1}{|p{1.2cm}|}{}                                & \multicolumn{1}{p{1.1cm}|}{\multirow{2}{*}{\parbox{1.1cm}{\raggedright With Answer}}}    & Time   & 7.82   & 11.84 & 55.90  \\ \cline{3-6} 
\multicolumn{1}{|p{1.2cm}|}{}                                & \multicolumn{1}{p{1.1cm}|}{}                                & Cost & 0.006  & 0.005 & 0.039  \\ \hline
\multicolumn{1}{|p{1.2cm}|}{\multirow{4}{*}{\parbox{1.2cm}{\raggedright GitHub}}}         & \multicolumn{1}{p{1.1cm}|}{\multirow{2}{*}{\parbox{1.1cm}{\raggedright Issues}}}         & Time   & 22.31  & 17.28 & 118.18 \\ \cline{3-6} 
\multicolumn{1}{|p{1.2cm}|}{}                                & \multicolumn{1}{p{1.1cm}|}{}                                & Cost & 0.008  & 0.014 & 0.025  \\ \cline{2-6} 
\multicolumn{1}{|p{1.2cm}|}{}                                & \multicolumn{1}{p{1.1cm}|}{\multirow{2}{*}{\parbox{1.1cm}{\raggedright Commits}}}        & Time   & 18.01  & 15.41 & 97.36  \\ \cline{3-6} 
\multicolumn{1}{|p{1.2cm}|}{}                                & \multicolumn{1}{p{1.1cm}|}{}                                & Cost  & 0.013  & 0.002 & 0.013  \\ \hline
\multicolumn{1}{|p{1.2cm}|}{\multirow{4}{*}{\parbox{1.2cm}{\raggedright Hugging Face}}}   & \multicolumn{1}{p{1.1cm}|}{\multirow{2}{*}{\parbox{1.1cm}{\raggedright Without Answer}}} & Time   & 20.19  & 25.78 & 100.38 \\ \cline{3-6} 
\multicolumn{1}{|p{1.2cm}|}{}                                & \multicolumn{1}{p{1.1cm}|}{}                                & Cost  & 0.021  & 0.024 & 0.041  \\ \cline{2-6} 
\multicolumn{1}{|p{1.2cm}|}{}                                & \multicolumn{1}{p{1.1cm}|}{\multirow{2}{*}{\parbox{1.1cm}{\raggedright With Answer}}}    & Time   & 9.39   & 14.69 & 75.67  \\ \cline{3-6} 
\multicolumn{1}{|p{1.2cm}|}{}                                & \multicolumn{1}{p{1.1cm}|}{}                                & Cost  & 0.007  & 0.010 & 0.015  \\ \hline
\end{tabular}
\end{table}

\vspace{-14pt}
\subsection{Efficiency and Cost}
\vspace{-5pt}
To evaluate the cost and efficiency of \name, we recorded the API cost and time taken per post or commit across the three platforms, as presented in Table \ref{tab:effectiveness_cost}. Claude Sonnet 4 consumed the most time and incurred the highest cost to reach a solution. Gemini 2.5 Flash required slightly more time than GPT o3 Mini on average, but unlike GPT o3 mini, it invoked multiple tools before arriving at a solution. When provided with a community answer, the agent made fewer tool calls and converged faster, thereby reducing the overall cost. We also conducted a pilot study with Claude Opus 4 on 110 Stack Overflow posts without accepted answers, which averaged 115.4 seconds and 0.455 USD per post.
\vspace{-1pt}

\vspace{-7pt}
\subsection{Comparison with Baselines}
\vspace{-1pt}
To compare~\name~with existing solutions, we evaluated four architectural settings covering both encoder-based and decoder-based LLMs.

\textbf{Encoder: } In this setting, input text is fed into an encoder transformer that generates a representation passed to a classification head. While fast and effective in specific cases \cite{hussain2025leveraging}, it requires labeled training data. We investigated CodeBERT-base \cite{feng2020codebert} and CodeT5-base \cite{wang2021codet5}, both known for strong performance in code-related tasks \cite{ghale2025automated}, fine-tuning on 80\% of the data and testing on the remaining 20\%.

\textbf{Zero-shot: } The LLM receives the input text and possible labels and selects one without additional information \cite{wang2023prompt}. This setting uses the same LLMs as \name.

\vspace{-2pt}

\textbf{One-shot:} Similar to zero-shot, but with one example per category. The shortest example by character length was selected to fit within the prompt window. For Stack Overflow and GitHub issues, even the shortest examples exceeded the contest window limit. To address this, the same LLM used for classification generated a concise summary of each example, which was included in the prompt instead of the full text. 

\textbf{Tool-less ReAct:} This setting mirrors~\name~but disables tool access. By returning a fixed ``No results found'' response whenever the agent attempts to invoke a tool, we establish a baseline to measure the performance gain specifically attributed to our tool integration.
\vspace{-5pt}
\begin{table}[h]
\centering
\caption{LLM comparison on Stack Overflow and GitHub tasks (W/O: Without). Complete table shared in \cite{AnonymousRepo12LLMAgentBugStudy}.}
\label{tab:comp}

\begin{adjustbox}{scale=0.79}
\begin{tabular}{|l|l|l|r|r|r|r|}
\hline
\multirow{2}{*}{\textbf{Approach}} & \multirow{2}{*}{\textbf{Model}} & \multirow{2}{*}{\textbf{Category}} 
& \multicolumn{2}{c|}{\textbf{Stack Overflow}} 
& \multicolumn{2}{c|}{\textbf{GitHub}} \\ \cline{4-7}

& & 
& \textbf{W/O Ans.} & \textbf{With Ans.} 
& \textbf{Commit} & \textbf{Issues} \\ \hline

\multirow{6}{*}{Encoder}
& \multirow{3}{*}{CodeBERT}
& Bug Type   & 0.19 & 0.29 & 0.41 & 0.33 \\ \cline{3-7}
& & Root Cause & 0.17 & 0.19 & 0.35 & 0.29 \\ \cline{3-7}
& & Effect     & 0.57 & 0.61 & 0.35 & 0.64 \\ \cline{2-7}

& \multirow{3}{*}{CodeT5}
& Bug Type   & 0.24 & 0.24 & 0.47 & 0.33 \\ \cline{3-7}
& & Root Cause & 0.17 & 0.10 & 0.40 & 0.23 \\ \cline{3-7}
& & Effect     & 0.55 & 0.58 & 0.36 & 0.64 \\ \hline

\multirow{3}{*}{Zero-shot}
& \multirow{3}{*}{Gemini}
& Bug Type   & 0.21 & 0.22 & 0.23 & 0.21 \\ \cline{3-7}
& & Root Cause & 0.13 & 0.18 & 0.11 & 0.24 \\ \cline{3-7}
& & Effect     & 0.35 & 0.29 & 0.18 & 0.40 \\ \hline

\multirow{3}{*}{One-shot}
& \multirow{3}{*}{Gemini}
& Bug Type   & 0.24 & 0.17 & 0.16 & 0.37 \\ \cline{3-7}
& & Root Cause & 0.18 & 0.18 & 0.14 & 0.22 \\ \cline{3-7}
& & Effect     & 0.50 & 0.56 & 0.11 & 0.37 \\ \hline

\multirow{3}{*}{\shortstack{Tool-less \\ ReAct}}
& \multirow{3}{*}{Gemini}
& Bug Type   & 0.30 & 0.20 & 0.19 & 0.30 \\ \cline{3-7}
& & Root Cause & 0.14 & 0.18 & 0.22 & 0.27 \\ \cline{3-7}
& & Effect     & 0.45 & 0.24 & 0.24 & 0.37 \\ \hline

\end{tabular}
\end{adjustbox}
\end{table}
\vspace{-12pt}

\subsection{Results Analysis and Discussion}
Table \ref{tab:comp} presents the F1-scores of all baseline architectures. For zero-shot, one-shot, and ReAct settings, only results of the best-performing LLM (Gemini 2.5 Flash) are shown for the critical components (bug type, root cause, and effect), with complete results shared in \cite{AnonymousRepo12LLMAgentBugStudy}. Among the three non-trivial categories,~\name~outperformed all baselines except in the effect category for GitHub commits, and outperformed most baselines on the trivial categories. The Tool-less ReAct agent achieved an average F1-score of only 25.71\% on non-trivial categories, while~\name~achieved 64.08\% using Gemini 2.5 Flash, demonstrating the significant impact of the developed tools. Encoder-based solutions also performed well. The one-shot approach underperformed zero-shot, likely because the example was selected based on length rather than category representativeness, which may have misled the LLM.

While~\name~outperformed all baselines, it has several limitations. The system depends on external sources, and their unavailability may significantly affect performance. Web scraping also introduces latency due to factors such as internet speed. Furthermore, the agent was evaluated on posts and small code snippets, and may require architectural adjustments to handle larger codebases. Overall,~\name~demonstrated strong performance in annotating LLM agent bugs. While prior studies reported limited self-healing ability in LLM agents \cite{rahardja2025can}, this work shows that agents can identify and annotate bugs in agentic systems when appropriately designed. Although~\name~targets bug annotation, it can be adapted for bug fixing by replacing the classification head with a fixation head, an adaptation future studies can explore.

\vspace{-5pt}

\section{Threats to Validity}
\label{sec:THREATSTOVALIDITY}

\textbf{Internal Threat: } One potential internal threat is bug classification. While no predefined taxonomy was used, some labels were adapted from prior literature and new ones were introduced where none existed. To mitigate the threat of potential bias in the annotation, two researchers independently annotated bugs and resolved disagreements through discussion with a knowledgeable moderator. We also provide rationales for each label across bug type, root cause, and effect. Another threat is the annotation of Stack Overflow and Hugging Face Forums posts without accepted answers. Posts with a verified solution were marked accordingly, and for others, external resources were referenced to support annotation decisions. Although both annotators verified these resources, minor subjectivity may remain since solutions were not confirmed by the users. Lastly, we acknowledge that the proposed taxonomy, though grounded in current literature and empirical observations, may evolve as LLM agent architectures advance.

\textbf{External Threat: } One potential threat is dataset inclusiveness, as libraries were selected based on popularity using GitHub stars \cite{awesome-llm-agents} and related Stack Overflow posts. Not all LLM agent libraries are represented, particularly those with limited adoption, and since the dataset is derived from community-reported bugs, popular frameworks are inherently overrepresented. Lastly,~\name~invokes 10 tools that scrape specific web pages, and any changes to these pages after April 2026 may affect tool calls and require adjustments to the scraping.

\vspace{-5pt}
\section{Conclusions and Future Work}
\label{sec:conc}
\vspace{-2pt}
In recent years, the use of LLM agents has grown significantly, while developers have faced critical challenges in building and maintaining these systems due to the novelty of the field and limited knowledge about the types of bugs that occur. To address this issue, we conducted the first comprehensive study of characterizing bugs in LLM agents. Our empirical study provides a deeper understanding of bugs across three community forums, namely Stack Overflow, GitHub, and Hugging Face discussions. Based on our findings, we provide actionable takeaways for developers, library maintainers, and researchers to improve practices and accommodate necessary changes. Additionally, we developed an LLM-based~\react~agent named~\name~to automatically annotate bugs from community platforms and evaluate the effectiveness of LLM agents in identifying and annotating bugs. Our study highlights the potential of LLMs, particularly Gemini 2.5 Flash, for this task at low cost. Future work should explore LLMs’ capabilities in fixing bugs in LLM agents and perform deeper analyses to achieve human-level performance in both annotating and resolving bugs.

\section*{Data Availability}
\vspace{-5pt}
Our benchmarks, results, findings, and implementation code of~\name~are available in this repository~\cite{AnonymousRepo12LLMAgentBugStudy}.
\vspace{-5pt}
\section*{Acknowledgment}
\vspace{-5pt}
This work was supported by the National Science Foundation (NSF) under Grant No. IIS-2419883. Any opinions, findings, and conclusions or recommendations expressed in this work are those of the author(s) and do not necessarily reflect the views of the NSF.
\vspace{-5pt}

\bibliographystyle{IEEEtran}
\bibliography{sample-base.bib}

\end{document}